\documentclass{aastex62}

\graphicspath{{./}{figures/}}

\usepackage{mathtools}
\usepackage{natbib}

\def\bbh#1{binary black hole#1
  (BBH#1)\gdef\bbh{BBH}}
  
\date{\today}

\shorttitle{LIGO's stellar budget}
\shortauthors{Jani \& Loeb}

\begin{document}

\title{Global Stellar Budget for LIGO Black Holes}

\correspondingauthor{Karan Jani}
\email{karan.jani@vanderbilt.edu}

\author[0000-0003-1007-8912]{Karan Jani}
\affil{Department Physics and Astronomy, Vanderbilt University, 2301 Vanderbilt Place, Nashville, TN, 37235, USA}

\author{Abraham Loeb}
\affil{Department of Astronomy, Harvard University, 60 Garden Street, Cambridge, MA 02138, USA}

\begin{abstract}
The binary black hole mergers observed by LIGO-Virgo gravitational-wave detectors pose two major challenges: (i) how to produce these massive black holes from stellar processes; and (ii) how to bring them close enough to merge within the age of the universe? We derive a fundamental constraint relating the binary separation and the available stellar budget in the universe to produce the observed black hole mergers. We find that $\lesssim 14\%$ of the entire budget contributes to the observed merger rate of $(30 + 30)~\mathrm{M}_\odot$ black holes, if the separation is around the diameter of their progenitor stars. Furthermore, the upgraded LIGO detector and third-generation gravitational-wave detectors are not expected to find stellar-mass black hole mergers at high redshifts. From LIGO's strong constraints on the mergers of black holes in the pair-instability mass-gap ($60-120~\mathrm{M}_\odot$), we find that $\lesssim 0.8\%$ of all massive stars contribute to a remnant black hole population in this gap. Our derived separation$-$budget constraint provides a robust framework for testing the formation scenarios of stellar binary black holes.
\end{abstract}


\section{Introduction} \label{sec:intro}
So far the network of LIGO-Virgo detectors have publicly announced 35 \bbh{} events, of which 10 are confirmed detections \citep{O2Catalog}. From this list of 30 confirmed black holes (primary, secondary, remnant), 80\% are much heavier than the black hole candidates found in X-ray binaries, thus suggesting a distinct evolutionary path for such binaries. Additionally, LIGO-Virgo has released a list of marginal triggers found in their different searches that may have astrophysical origin but cannot be confirmed due to their relatively low detection statistics. 

Among this rich gravitational-wave data set lie two exceptional cases that hint at a completely new population of black holes: (i) GW170729, whose primary black hole mass lies in the range $[40.4-67.2]~\mathrm{M}_\odot$ with 90\% confidence, and (ii) 170502, the loudest marginal trigger published so far by LIGO-Virgo \citep{O2IMBH}, which has a total-mass in the range $ [123-239]~\mathrm{M}_\odot$. Assuming this trigger is astrophysical, and of equal-mass (like other LIGO \bbh{s}), we get a primary black hole in the range $[61-120]~\mathrm{M}_\odot$. Both GW170729 and 170502 offer new evidence of black holes in the pair-instability gap range $(60-120~\mathrm{M}_\odot)$ - a mass-interval forbidden for supernova remnants \citep{Woosley_2017}. 

Observations of these black holes have further complicated the question of how massive stars end up as tight \bbh{} systems. A non-exhaustive list of proposed scenarios to tackle this includes mass-exchange in binary stars \citep{2016Natur.534..512B, de_Mink_2016}, multiple mergers of young stars \citep{Di_Carlo_2019}, dynamical segregation of black holes in star clusters \citep{Rodriguez_2016} and binary formation within a single rapidly rotating star \citep{D_Orazio_2018}. While each scenario has a distinct observable signature (involving spin orientations, mass-ratios) and predicted merger rate, they ought to be fundamentally constrained by the number of stars in the universe. We take this as a starting point and derive a global budget for \bbh{} mergers. We show that regardless of the proposed scenario, the progenitor star budget already imposes a strict limit on the \bbh{} separation.

  \section{Methodology} \label{sec:methods}

\begin{figure*}
\centering
 \includegraphics[scale=0.9,trim = {60 130 30 50}]{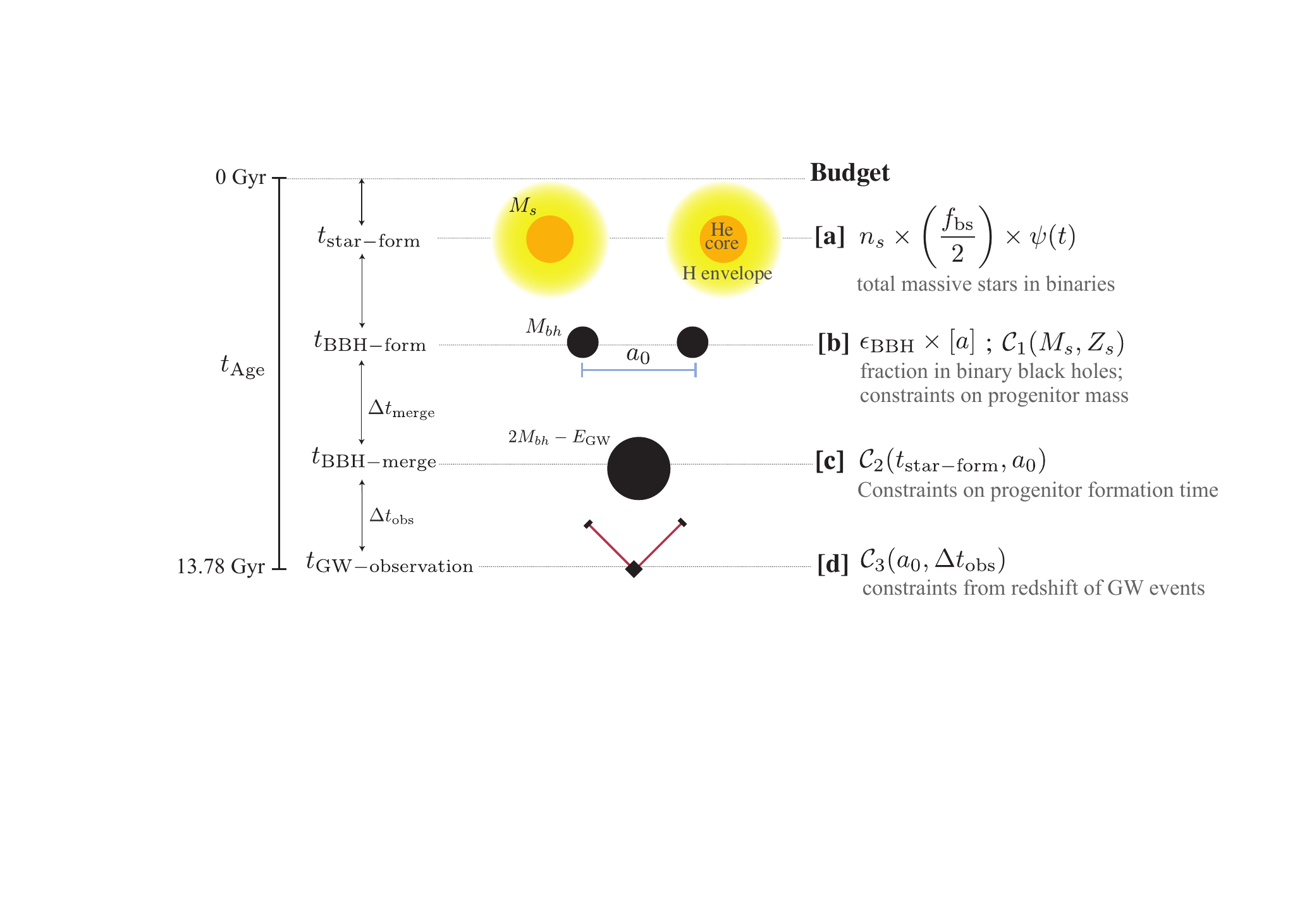}
\caption{{ Diagram of stellar budget for merging black holes at different snapshots of time. }
}
\label{fig1} 
\end{figure*}

\begin{figure*}
\centering
 \includegraphics[scale=0.75,trim = {40 50 30 0}]{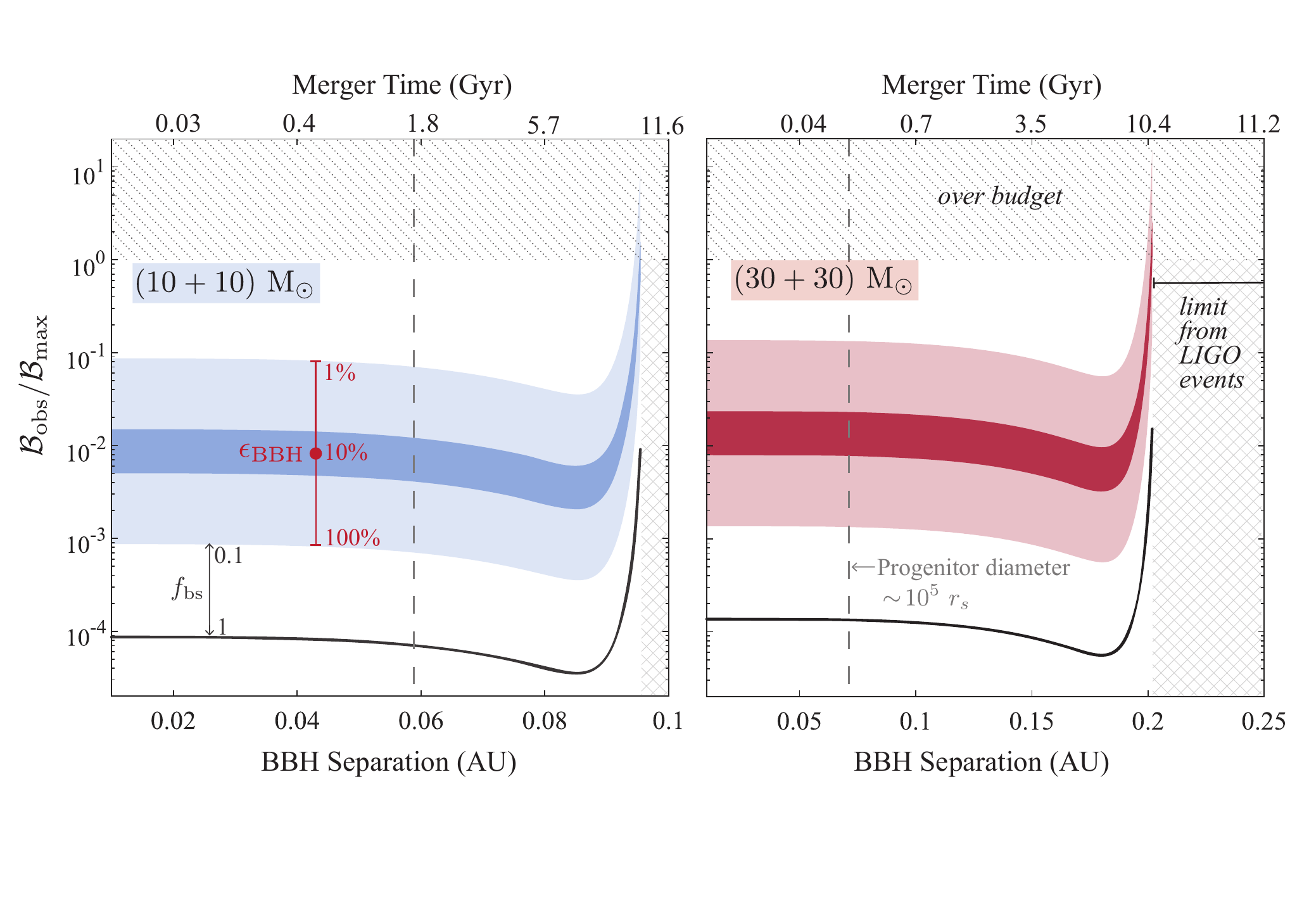}
  \includegraphics[scale=0.75,trim = {40 60 30 20}]{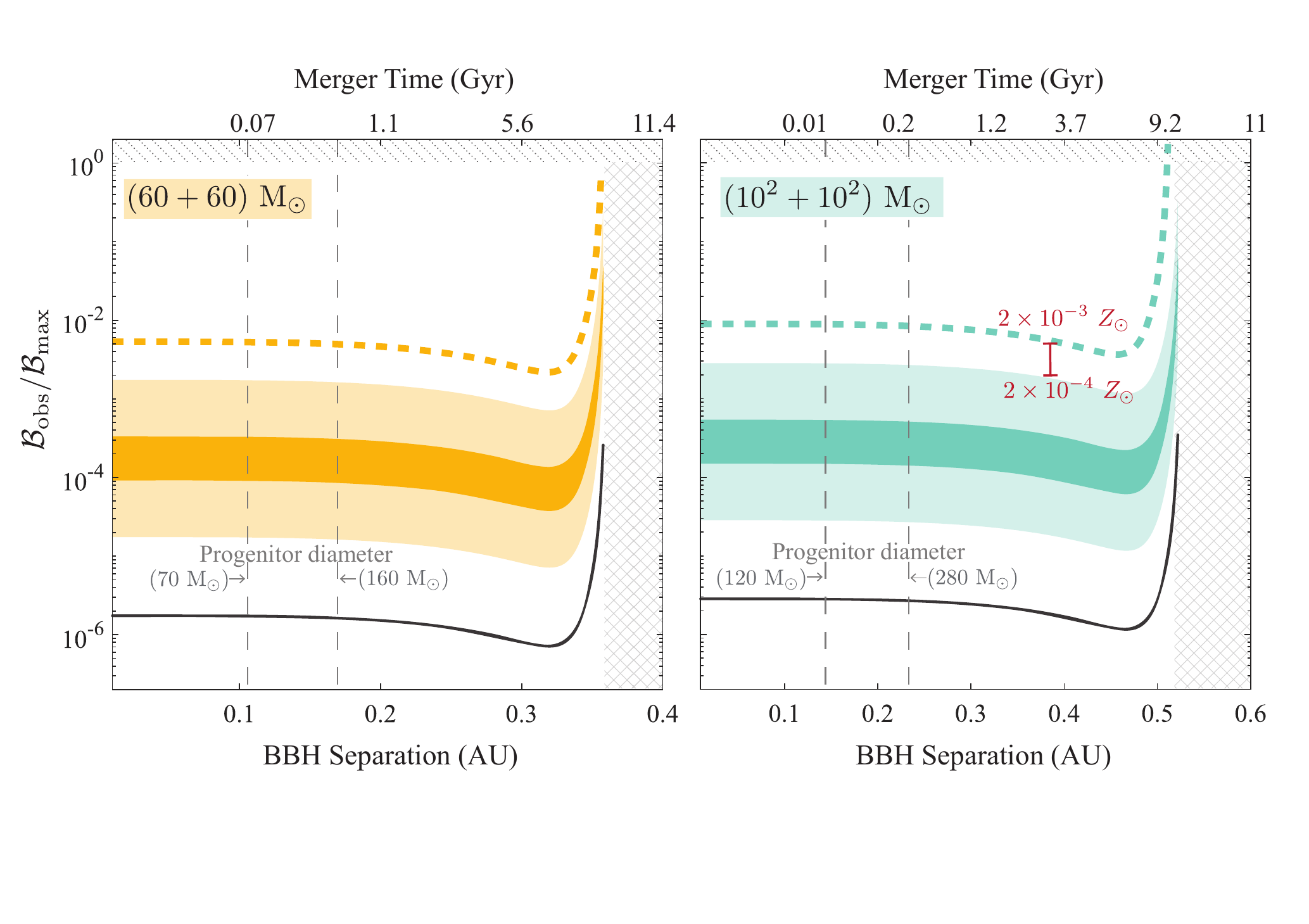}
\caption{{ Progenitor star budget for different LIGO-Virgo \bbh{s} as a function of initial separation. } The vertical axis shows the observed fraction  $\mathcal{B}_{obs}/\mathcal{B}_{max}$ of progenitor remnants. {\it Top panel:} stellar \bbh{} currently observed by LIGO. {\it Bottom panel:} binaries of black holes in the pair-instability mass-gap. The colored patches refers to the assumed stars-to-\bbh{} efficiency parameter, $\epsilon_\mathrm{BBH}$. The thick line in their center corresponds to $\epsilon_\mathrm{BBH}=0.1$, i.e. 10\% of all massive stars turn into LIGO black holes. The black curve in each case refers to the theoretical maximum $(\epsilon_\mathrm{BBH}=1;~f_\mathrm{bs}=1$). The thick dotted line in the bottom panel shows the effect of lower metallicity. The dashed area from the top highlights the over budget ($\mathcal{B}_{obs}/\mathcal{B}_{max}>1)$, and the dashed area from the right shows the constraint on the initial \bbh{} separation, $a_0$, from the average redshift of LIGO events. 
}
\label{fig2} 
\end{figure*}

As shown diagrammatically in Figure \ref{fig1}, the maximum budget for producing mergers of \bbh{s} from stellar evolution can be calculated as, 

\begin{equation}
      \mathcal{B}_{max} = 
    \underbrace{{n}_s(M_s) \times ~\left( \frac{f_\mathrm{bs}}{2} \right)}_\text{Massive stars in binary}~ \times 
       \underbrace{ \epsilon_\mathrm{BBH} }_\text{Formation channel efficiency}  \times \underbrace{\psi(z_\mathrm{star-form}) \times \left( \frac{1+z_\mathrm{BBH-merge}}{1+z_\mathrm{star-form}} \right)}_\text{Comoving star formation rate}
\label{eq1}        
\end{equation}

\noindent with three astrophysical constraints, 

\begin{equation}
    \underbrace{\mathcal{C}_1 \left( {M_s}/{\mathrm{M}_\odot}  \leftrightarrow	 {Z_{s}}/{{Z}_\odot} \right)}_\text{Mass of progenitor from metalicity}~, \hspace{0.3in}
        \underbrace{\mathcal{C}_2({z_\mathrm{star-form}} \leftrightarrow	 a_0) }_\text{Birth rate of progenitor from separation }, \hspace{0.3in}
        \underbrace{\mathcal{C}_3( \mathrm{max}{\{a_0\}}  \leftrightarrow	 {z_\mathrm{obs}})}_\text{Separation from GW detection} 
\label{eq2}        
\end{equation}

\noindent Here,
\begin{equation}
      n_s = N_{>M_s} / M_{tot} =  {\int_{M_s}^{M_{max}}{\xi(m) ~\mathrm{d}m}} ~/~ {\int_{0.01~\mathrm{M}_\odot}^{1000~\mathrm{M}_\odot} m~\xi(m) ~\mathrm{d}m}, 
\label{eq3}        
\end{equation}

\noindent is the number of progenitor stars per unit stellar mass that will result as black holes.  We utilize the piecewise initial-stellar mass function $\xi(m)$ of \cite{Kroupa_2001} and include the numerical treatment that finds maximum star mass $M_{max}$, given a fundamental cutoff at $1000~\mathrm{M}_\odot$ \citep{2005IAUS..227..423K}. The choice for progenitor mass $M_s$ is determined by the desired black hole mass $M_\mathrm{BH}$. We use results from the population synthesis code of \cite{Spera_2017} to find the mapping between the mass of progenitor star and remnant black hole (for simplicity, excluding the treatment of pair-instability mass-gap on $M_{BH}$). For the massive stars that produce LIGO black holes, this mapping can vary significantly with the assumed metallicity, ${Z}/{Z}_\odot$, of the progenitor stars, $t_\mathrm{star-form}$. This dependence is related to the assumed \bbh{} formation channel, and sets a relation, $\mathcal{C}_1$, between metallicity and the number of massive stars. 

The fraction of progenitor stars in a tight binary is set by the parameter $f_\mathrm{bs} \in [0,1]$. While $70\%$ of O-stars within the Milky Way are essentially binaries, this fraction is lower for the ones that are tightly bounded with orbital period of days \citep{Sana_2012}. Such binaries provide a favorable chance for LIGO black holes. Therefore, unless stated otherwise, we adopt $f_\mathrm{bs} = 0.1$ throughout this study. 

The efficiency for converting binaries of massive stars into gravitationally bound \bbh{} sources is captured by $\epsilon_\mathrm{BBH} \in [0,1]$. This free parameter solely depends on the assumed formation channel. From the asymmetric collapse, the remnant formed as black hole can get a natal kick \citep{Hoogerwerf_2001}, thus decreasing the fraction ($\epsilon_\mathrm{BBH}$) of \bbh{} from two progenitor massive stars. If LIGO's \bbh{s} are formed through dynamical capture \citep{Rodriguez_2016}, then $\epsilon_\mathrm{BBH}$ is the fraction of black holes that will find a similar partner. 

From the LIGO observations, we cannot infer the distribution of initial separation $a_0$ at $t_\mathrm{BBH-form}$, the instance when the two black holes become gravitationally bound. Therefore, we adopt a delta function for a given choice of the initial \bbh{} separation $a_0$ (AU). Assuming a circular orbit and absence of any external influence, the separation $a_0$ sets a bound on $\Delta{t}_\mathrm{merge}$ \citep{Peters1964}. As the time from the birth to collapse, $\Delta{t}_\mathrm{col} \ll\Delta{t}_\mathrm{merge}$, we can assume that the progenitor stars were formed at $t_\mathrm{star-form}\sim t_\mathrm{BBH-form}$.  The number of progenitor stars in the universe at this instance can be found through stellar formation rate, $\psi(z)$ [see equation (16) of \cite{Madau_2014}]. This sets the second constraint, $\mathcal{C}_2$, which relates $a_0$ with the production of progenitor stars. As we are using cosmic star formation rate at an earlier epoch, $(1 + z_\mathrm{BBH-form})$, to calculate the LIGO event rate at $(1+z_\mathrm{BBH-merge})$, there will be a time lag for the corresponding $\psi$. Furthermore, the observed redshift $z_\mathrm{BBH-merge}$ of LIGO detections provides a constraint, $\mathcal{C}_3$, on the maximum value of $a_0$ such that the merger time, $\Delta{t}_\mathrm{merge}$, is less than the Hubble time at that redshift.

\section{Results and Discussion} \label{sec:results}
Based on equations (\ref{eq1})$-$(\ref{eq3}), we can compute the fraction of stellar budget that is being utilized for $\mathcal{B}_{obs}$ - the observed merger on \bbh{} from LIGO. If $ \mathcal{B}_{obs}/\mathcal{B}_{max}>1$, then there are not enough stars in the universe to produce these merging black holes. This global budget calculation allows us to put fundamental limits on $a_0$, regardless of the mechanism that brings them close enough. Figure \ref{fig2} shows the fraction of the total allowed budget being utilized for four distinct populations of \bbh{} mergers that are strongly constrained by LIGO observations. 

\paragraph{Stellar-mass \bbh{s} } To produce black holes $10-30~\mathrm{M}_\odot$, the typical mass of the progenitor star ranges from $25-35~\mathrm{M}_\odot$. For metallicities $\lesssim 10^{-3}~Z_\odot$, this mass-range remains fairly unaffected. Thus, the constraint $\mathcal{C}_1$ can be ignored for the vanilla \bbh{} events detected in LIGO. We take the upper-limit $\mathcal{B}_{obs} = 111.7~\mathrm{Gpc}^{-3} ~\mathrm{yr}^{-1}$ across the stellar \bbh{} mass-range $\lesssim 100~\mathrm{M}_\odot$ \citep{O2Rates}. 

For a $(30+30)~\mathrm{M}_\odot$ source (such as GW150914), we find the stellar budget sets stringent constraints on binary separation. Adopting $a_0=0.2$ AU can max out the entire stellar budget (see top-right panel of Figure \ref{fig2}). If these \bbh{s} are separated by the diameter of one of their progenitor star of $35~\mathrm{M}_\odot$ \citep{1991Ap&SS.181..313D}, which corresponds to $\sim 10^5$ Schwarzschild radii $(r_s)$, the progenitor star budget utilized would be $\lesssim14\%$ (at low efficiency, $\epsilon_\mathrm{BBH}=0.01$). The budget remains fairly constant for any lower separation from progenitor's diameter. Furthermore, we find a tipping point at $a_0=0.18$ AU, which assures the lowest stellar budget is being utilized. This separation leads to $t_\mathrm{BBH-form}$ at the peak of star formation.   

For a $(10+10)~\mathrm{M}_\odot$ source (such as GW151226), we find that a binary separation of just $a_0=0.094$ AU can max out the entire progenitor star budget. If these \bbh{s} are separated by the diameter of one of their $25~\mathrm{M}_\odot$ progenitor star, the  budget utilized would be $\lesssim7\%$ (at low efficiency, $\epsilon_\mathrm{BBH}=0.01$). The tipping point for these \bbh{} masses happen at $a_0=0.085$ AU.

\paragraph{Pair-Instability \bbh{s} } From the detection threshold set by the trigger 170502, \cite{O2IMBH} provided upper-limits on the population of merging black holes in the pair-instability gap. From their published list, we take two cases $(60+60)~\mathrm{M}_\odot$ and $(100+100)~\mathrm{M}_\odot$, which has $\mathcal{B}_{obs} = 0.56$ and $0.44~\mathrm{Gpc}^{-3}$ respectively. Because their progenitor mass can vary significantly with the assumed metallicity, the constraint $\mathcal{C}_1$ becomes important in determining their budget. 

For these black holes, we note an interesting result - at no binary separation they tend to exceed the stellar budget (see bottom panel of Figure \ref{fig2}). When separated by the diameter of one of their progenitor star, $(100+100)~\mathrm{M}_\odot$ would at most utilize 0.8\% (0.3\%) budget for metalicity of $2\times10^{-3}~{Z}_\odot$ ($2\times10^{-4}~{Z}_\odot$). The maximum budget is spent when $a_0 = 0.52$ AU. At this separation, high metalicity of progenitor star may just max out the budget, but it is unlikely for stars that early in the universe.  The tipping point for these \bbh{} masses happen at $a_0=0.46$~AU. In the case of $(60+60)~\mathrm{M}_\odot$ \bbh{} merger, the budget utilized would be  0.2\% (for $2\times10^{-3}~{Z}_\odot$). The maximum budget for such \bbh{s} is spent when $a_0 = 0.36$ AU. From this analysis, we conclude that more confident detections of pair-instability black holes by LIGO would likely correspond to rare progenitors whose population is at least an order of magnitude less than the entire progenitor population. This constraint is consistent with the expectation of no \bbh{} in this mass-range based on stellar evolution \citep{Woosley_2017}.

\begin{figure*}
\centering
 \includegraphics[scale=0.85,trim = {40 50 30 0}]{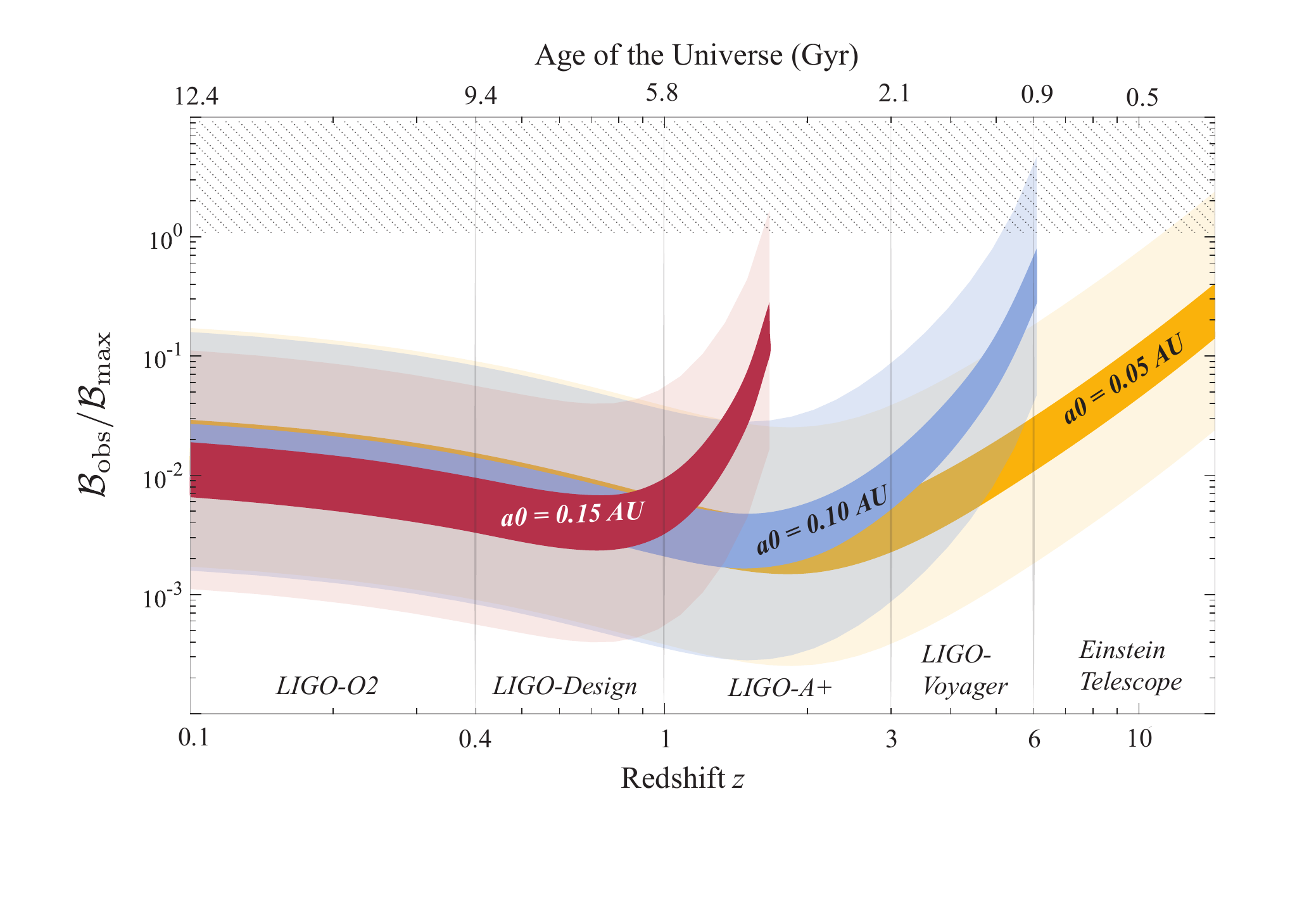}
\caption{{ Progenitor star budget for $(30+30)~\mathrm{M}_\odot$ events as a function of the observed redshift and the corresponding age of the universe (upper horizontal axis label). }  The vertical axis $\mathcal{B}_{obs}/\mathcal{B}_{max}$ with the colored patches refer to three values of binary separation $a_0$. Like Figure \ref{fig2}, their width refers to the assumed efficiency parameter $\epsilon_\mathrm{BBH}$. The thick line in their center is for $\epsilon_\mathrm{BBH}=0.1$. The hashed area near the top highlights the over budget ($\mathcal{B}_{obs}/\mathcal{B}_{max}>1)$. The redshift has been highlighted for different epochs of gravitational-wave experiments based on their sensitivity for a $(30+30)~\mathrm{M}_\odot$  \citep{jani2019detectability}. }
\label{fig3} 
\end{figure*}

\paragraph{Future Detectors} As the sensitivity improves, LIGO-like detectors can detect \bbh{} mergers from earlier cosmic times \citep{ObservingScenarios}. For mergers of  $(30+30)~\mathrm{M}_\odot$, the upgraded versions of LIGO (mid-2020s) would permit detection up to redshift $z\sim6$, while third-generation detectors in 2030s such as Einstein Telescope \citep{Punturo:2010zz} and Cosmic Explorer \citep{2019arXiv190704833R} can find these events up to $z\lesssim 40$ \citep{jani2019detectability}. If mergers are found at high redshift, the constraint $\mathcal{C}_3$ on maximum $a_0$ becomes even more stringent. Figure \ref{fig3} shows the budget that would be utilized for different choices of $a_0$ as a function of detection redshift. If \bbh{s} are separated initially at $0.1$~AU ($0.15$~AU), then their detection by redshift $5$ $(1.5)$ would max out the entire budget. We can start constraining $\epsilon_\mathrm{BBH}$ if the current LIGO facility sees mergers at $z\gtrsim1$. The next-generation detectors would constraint $a_0$ to the lowest realistic values.   

\paragraph{Conclusion} We derived global constraints on binary black hole separation and the stellar budget of their progenitor stars. In particular, we find that at most $14\%$ of the available cosmic budget contributes to the observed merger rate of $(30+30)~\mathrm{M}_\odot$ black holes. For the black hole mergers accessible to LIGO, we find a general trend, where up to a tipping point $a_0$ dictated by their mass, the progenitor star budget remains fairly constant, and thereafter gets maxed out exponentially. Our results indicate that observations beyond redshift $\gtrsim1$ of $(30+30)~\mathrm{M}_\odot$ \bbh{s} - the most common source for current ground-based gravitational-wave astronomy - can strongly constrain the efficiency ($\epsilon_\mathrm{BBH}$) of converting stars into merging black holes. These global considerations provides a model independent framework for testing all possible formation channels for \bbh{s}.

\acknowledgments
We thank Kelly Holley-Bockelmann for helpful discussions. K.J's research was supported by the GRAVITY program at Vanderbilt University. This work was supported in part by the Black Hole Initiative at Harvard University, which is funded by grants from the John Templeton Foundation and the Gordon and Betty Moore Foundation.

\bibliographystyle{aasjournal}

\bibliography{references}

\begin{thebibliography}{}
\expandafter\ifx\csname natexlab\endcsname\relax\def\natexlab#1{#1}\fi
\providecommand{\url}[1]{\href{#1}{#1}}

\bibitem[{{Belczynski} {et~al.}(2016){Belczynski}, {Holz}, {Bulik}, \&
  {O'Shaughnessy}}]{2016Natur.534..512B}
{Belczynski}, K., {Holz}, D.~E., {Bulik}, T., \& {O'Shaughnessy}, R. 2016,
  Nature, 534, 512

\bibitem[{de~Mink \& Mandel(2016)}]{de_Mink_2016}
de~Mink, S.~E., \& Mandel, I. 2016, Monthly Notices of the Royal Astronomical
  Society, 460, 3545–3553.
\newblock \url{http://dx.doi.org/10.1093/mnras/stw1219}

\bibitem[{{Demircan} \& {Kahraman}(1991)}]{1991Ap&SS.181..313D}
{Demircan}, O., \& {Kahraman}, G. 1991, \apss, 181, 313

\bibitem[{Di~Carlo {et~al.}(2019)Di~Carlo, Giacobbo, Mapelli, Pasquato, Spera,
  Wang, \& Haardt}]{Di_Carlo_2019}
Di~Carlo, U.~N., Giacobbo, N., Mapelli, M., {et~al.} 2019, Monthly Notices of
  the Royal Astronomical Society, 487, 2947–2960.
\newblock \url{http://dx.doi.org/10.1093/mnras/stz1453}

\bibitem[{D’Orazio \& Loeb(2018)}]{D_Orazio_2018}
D’Orazio, D.~J., \& Loeb, A. 2018, Physical Review D, 97,
  doi:10.1103/physrevd.97.083008.
\newblock \url{http://dx.doi.org/10.1103/PhysRevD.97.083008}

\bibitem[{Hoogerwerf {et~al.}(2001)Hoogerwerf, de~Bruijne, \&
  de~Zeeuw}]{Hoogerwerf_2001}
Hoogerwerf, R., de~Bruijne, J. H.~J., \& de~Zeeuw, P.~T. 2001, Astronomy \&
  Astrophysics, 365, 49–77.
\newblock \url{http://dx.doi.org/10.1051/0004-6361:20000014}

\bibitem[{Jani {et~al.}(2019)Jani, Shoemaker, \&
  Cutler}]{jani2019detectability}
Jani, K., Shoemaker, D., \& Cutler, C. 2019, Detectability of Intermediate-Mass
  Black Holes in Multiband Gravitational Wave Astronomy, , , arXiv:1908.04985

\bibitem[{Kroupa(2001)}]{Kroupa_2001}
Kroupa, P. 2001, Monthly Notices of the Royal Astronomical Society, 322,
  231–246.
\newblock \url{http://dx.doi.org/10.1046/j.1365-8711.2001.04022.x}

\bibitem[{{Kroupa} \& {Weidner}(2005)}]{2005IAUS..227..423K}
{Kroupa}, P., \& {Weidner}, C. 2005, in IAU Symposium, Vol. 227, Massive Star
  Birth: A Crossroads of Astrophysics, ed. R.~{Cesaroni}, M.~{Felli},
  E.~{Churchwell}, \& M.~{Walmsley}, 423--433

\bibitem[{Madau \& Dickinson(2014)}]{Madau_2014}
Madau, P., \& Dickinson, M. 2014, Annual Review of Astronomy and Astrophysics,
  52, 415–486.
\newblock \url{http://dx.doi.org/10.1146/annurev-astro-081811-125615}

\bibitem[{{Peters}(1964)}]{Peters1964}
{Peters}, P.~C. 1964, Physical Review, 136, 1224

\bibitem[{Punturo {et~al.}(2010)}]{Punturo:2010zz}
Punturo, M., {et~al.} 2010, Class. Quant. Grav., 27, 194002

\bibitem[{{Reitze} {et~al.}(2019){Reitze}, {Adhikari}, {Ballmer}, {Barish},
  {Barsotti}, {Billingsley}, {Brown}, {Chen}, {Coyne}, {Eisenstein}, {Evans},
  {Fritschel}, {Hall}, {Lazzarini}, {Lovelace}, {Read}, {Sathyaprakash},
  {Shoemaker}, {Smith}, {Torrie}, {Vitale}, {Weiss}, {Wipf}, \&
  {Zucker}}]{2019arXiv190704833R}
{Reitze}, D., {Adhikari}, R.~X., {Ballmer}, S., {et~al.} 2019, arXiv e-prints,
  arXiv:1907.04833

\bibitem[{Rodriguez {et~al.}(2016)Rodriguez, Haster, Chatterjee, Kalogera, \&
  Rasio}]{Rodriguez_2016}
Rodriguez, C.~L., Haster, C.-J., Chatterjee, S., Kalogera, V., \& Rasio, F.~A.
  2016, The Astrophysical Journal, 824, L8.
\newblock \url{http://dx.doi.org/10.3847/2041-8205/824/1/L8}

\bibitem[{Sana {et~al.}(2012)Sana, de~Mink, de~Koter, Langer, Evans, Gieles,
  Gosset, Izzard, Le~Bouquin, \& Schneider}]{Sana_2012}
Sana, H., de~Mink, S.~E., de~Koter, A., {et~al.} 2012, Science, 337, 444–446.
\newblock \url{http://dx.doi.org/10.1126/science.1223344}

\bibitem[{Spera \& Mapelli(2017)}]{Spera_2017}
Spera, M., \& Mapelli, M. 2017, Monthly Notices of the Royal Astronomical
  Society, 470, 4739–4749.
\newblock \url{http://dx.doi.org/10.1093/mnras/stx1576}

\bibitem[{{The LIGO Scientific Collaboration} \& {the Virgo
  Collaboration}(2013)}]{ObservingScenarios}
{The LIGO Scientific Collaboration}, \& {the Virgo Collaboration}. 2013,
  arXiv:1304.0670, arXiv:1304.0670

\bibitem[{{The LIGO Scientific Collaboration} \& {the Virgo
  Collaboration}(2018)}]{O2Catalog}
---. 2018, arXiv e-prints, arXiv:1811.12907

\bibitem[{{The LIGO Scientific Collaboration} \& {the Virgo
  Collaboration}(2019)}]{O2IMBH}
---. 2019, Physical Review D, 100, doi:10.1103/physrevd.100.064064.
\newblock \url{http://dx.doi.org/10.1103/PhysRevD.100.064064}

\bibitem[{{The LIGO Scientific Collaboration} \& {The Virgo
  Collaboration}(2019)}]{O2Rates}
{The LIGO Scientific Collaboration}, \& {The Virgo Collaboration}. 2019, The
  Astrophysical Journal, 882, L24.
\newblock \url{https://doi.org/10.3847\%2F2041-8213\%2Fab3800}

\bibitem[{Woosley(2017)}]{Woosley_2017}
Woosley, S.~E. 2017, The Astrophysical Journal, 836, 244.
\newblock \url{http://dx.doi.org/10.3847/1538-4357/836/2/244}

\end{thebibliography}



\end{document}